\documentclass[onecolumn,prd,aps,floats,nofootinbib,showpacs]{revtex4}
\usepackage{epsfig}
\usepackage[dvips]{color}

\newcommand{\be}{\begin{equation}}
\newcommand{\ee}{\end{equation}}
\newcommand{\bea}{\begin{eqnarray}}
\newcommand{\eea}{\end{eqnarray}}

\def\bu{{\mathbf u}}

\def\Ep{E^\prime}
\def\Ip{I^\prime}
\def\d{{\rm d}}
\def\sigv{\langle\sigma_{\rm ann}v\rangle}
\def\alt{\raise0.3ex\hbox{$\;<$\kern-0.75em\raise-1.1ex\hbox{$\sim\;$}}}
\def\agt{\raise0.3ex\hbox{$\;>$\kern-0.75em\raise-1.1ex\hbox{$\sim\;$}}}
\begin{document}

\title{Angular Signatures of Dark Matter in the Diffuse Gamma Ray Background}
\author{Dan Hooper and Pasquale D. Serpico}
\affiliation{Center for Particle Astrophysics, Fermi National
Accelerator Laboratory, Batavia, IL 60510-0500, USA}
\begin{abstract}

Dark matter annihilating in our Galaxy's halo and elsewhere in the
universe is expected to generate a diffuse flux of gamma rays,
potentially observable with next generation satellite-based
experiments, such as GLAST. In this article, we study the signatures
of dark matter in the angular distribution of this radiation, in
particular the deterministic ones. Pertaining to the extragalactic
contribution, we discuss the effect of the motion of the solar
system with respect to the cosmological rest frame, and anisotropies
due to the structure of our local universe. For the gamma ray flux
from dark matter in our own Galactic halo, we discuss the effects of
the offset position of the solar system, the Compton-Getting effect,
the asphericity of the Milky Way halo, and the signatures of nearby
substructure. We explore the prospects for the detection of these
features by the GLAST satellite and find that, if $\sim$10\% or more
of the diffuse gamma ray background observed by EGRET is the result
of dark matter annihilations, then GLAST should be sensitive to
anisotropies at the sub-percent level. Such precision would be
sufficient to detect many, if not all, of the signatures discussed
in this paper.

\end{abstract}
\pacs{95.35.+d, 
95.85.Pw, 
98.70.Vc    
\hfill FERMILAB-PUB-07-038-A}

\maketitle
\section{Introduction}

The spectrum of gamma rays generated in dark matter annihilations
has the potential to serve as a valuable tool in exploring the
physical and astrophysical properties of dark matter. As the dark
matter annihilation rate scales with the square of the density, one
would naively expect the brightest sources of gamma rays to be
nearby and high-density objects, such as the Galactic
Center~\cite{gc}, dwarf spheroidal galaxies~\cite{dwarf}, or other
local dark matter structures~\cite{sub}.

It is not necessarily the case, however, that these sources are the
easiest targets to detect the flux of dark matter annihilation
radiation. In particular, the distribution of dark matter in the
Galactic Center is not well known. Although N-body simulations
predict that dark matter halos should contain high density cusps in
their centers~\cite{NFW,Moore}, mechanisms have been proposed which
would modify this prediction. This is particularly true in the
innermost regions of these halos, where additional phenomena
involving baryons or central black holes may be
relevant~\cite{Merritt:2006mt}. As the rate of dark matter
annihilations in these halos depends critically on the density in
their innermost centers, it is extremely difficult to reliably
estimate the brightness of dark matter annihilation radiation from
this class of regions. Although the profiles of dwarf spheroidals
are better constrained, to a minor extent these targets suffer of
similar problems.

Astrophysical sources of gamma rays can also provide a formidable
background for dark matter searches from local, high density
sources. In particular, the Galactic Center has been found by
H.E.S.S. and other gamma ray telescopes to contain a bright source
of very high energy radiation~\cite{gctev}. The presence of this
source poses a very serious challenge to future dark matter searches
in the Galactic Center~\cite{Zaharijas:2006qb}.

With these issues in mind, it is difficult to assess the expected
signal and corresponding background expected from annihilating dark
matter in nearby, high density regions, even within the context of a
well-defined particle physics model. Moreover, even if such a source
were observed, one might be concerned whether it could be reliably
identified as dark matter radiation, rather than as another class of
astrophysical object. Only with a high precision measurement of the
gamma ray spectrum, ideally including the detection of
mono-energetic lines, could point sources of dark matter
annihilation radiation be conclusively identified
\cite{Baltz:2006sv}.

An alternative target for indirect dark matter searches is the
diffuse, or unresolved, contribution to the gamma ray
spectrum~\cite{Bergstrom:2001jj,Ullio:2002pj,Elsaesser:2004ck}.
Cosmological data---in particular the cosmic microwave background
anisotropies and the observations of the large scale matter power
spectrum---strongly favor (relatively) cold dark matter. In turn,
this implies that structure formation takes place in a hierarchical
manner, i.e. small objects formed at earlier times and larger
objects formed at later times from their subsequent merging. As a
result of this process, dark matter structures (``halos'') are
expected to host large numbers of clumps (``sub-halos''), which are
the relics of the formation history. Given the fact that the
annihilation signal scales with the square of the dark matter
density, the presence of substructure can result in an enhancement
of several orders of magnitude in the gamma ray flux with respect to
the naive estimate for a smooth distribution, as first noticed
in~\cite{Silk:1992bh}. For a simple argument justifying this
conclusion, see Appendix~\ref{simpleargument}.

The gamma ray signal from dark matter halos (including our own
Galaxy) depends crucially on the fraction of sub-halos that survives
the hierarchical merging process, as well as their density profile
and spatial distribution. The lack of knowledge of the dark matter
power spectrum on very small scales, the difficulty involved in
calculating gravitational clustering in the deep non-linear regime,
and uncertainties in the effects of the tidal stripping of sub-halos
by both dark matter and baryonic structures conspire to make current
estimates of the average annihilation rate uncertain by orders of
magnitude, independently of unknowns regarding the particle identity
of the dark matter candidate. At present, it is even unclear whether
the diffuse dark matter signal should be dominated by our Galactic
halo or by the extragalactic component. Furthermore, even if our
Galaxy's halo dominates the diffuse gamma ray spectrum, it is not
clear from which direction, and to what extent, the emission from
dark matter sub-halos will dominate the overall flux from dark
matter annihilations.

A lot of attention has been given in the literature to studying the
spectrum of gamma rays in order to identify signatures of dark
matter (for a recent example, see Ref.~\cite{Baltz:2006sv}).
Although this has the potential to be a powerful diagnostic tool, it
will be challenging to clearly separate a dark matter signal from
alternative astrophysical sources unless very precise measurements
are performed with a larger number of events and/or over a large
energy range. This is especially true in the case of diffuse
radiation and in scenarios where dark matter annihilations only
provide a subleading contribution to the total gamma ray flux.

In this paper, we mainly focus our attention to the complementary
information provided by the angular patterns in the diffuse
radiation generated in dark matter annihilations, stressing in
particular on deterministic features. In
Sec.~\ref{galexgal}, we provide a simple parametrization of the
Galactic and extragalactic diffuse dark matter fluxes and discuss
under which circumstances each of these components dominate.
Secondly, we summarize the key signatures and distinctive features
of the dark matter diffuse gamma ray radiation if dominated by the
extragalactic (Sec.~\ref{extragal}) or Galactic contribution
(Sec.~\ref{gal}). We also discuss the prospects for the detection of
these signatures by the forthcoming generation of instruments, in
particular the GLAST satellite (Sec.~\ref{detection}). We summarize
our conclusions in Sec.~\ref{summary}.

\section{Galactic vs. extragalactic dominance}\label{galexgal}
In this section, we compare the relative importance of the Galactic
and extragalactic contributions to the diffuse gamma ray emission.
Depending on which of these dominates, different signatures are
expected in future gamma ray observations. Although the average dark
matter density in both cases is pretty well known, because of the
effect of substructures discussed in Sec. I and illustrated in
Appendix~\ref{simpleargument}, the very prediction of the {\it
average} flux is a highly non-trivial problem. In general, apart for
particle physics details, it depends on several astrophysical and
cosmological parameters. In the following, we shall introduce simple
phenomenological models which condense the main uncertainties in a
limited number of unknowns, in order to allow for a simple
comparison of the two fluxes.

We begin by computing the galactic halo emission. This is made by
two contributions: one from the smooth halo and another by the
fraction of the density retained in sub-halos. We closely follow the
approach of Ref.~\cite{Berezinsky:2003vn}, although we retain the
angular and energy dependence and use a slightly different notation.
The differential flux of photons (in units of photons per area per
time per steradian per energy) from dark matter annihilations with a
smooth distribution can be written as
\begin{equation}
I_{\rm sm}(E,\psi)=f_{\gamma}(E)\,\Pi\,\int_{\rm l.o.s.} \d s\,\frac{\rho_{\rm sm}^2[r(s,\psi)]}{4\pi}, \label{Ism}
\end{equation}
where
\begin{equation}
r(s,\psi)=\sqrt{r_\odot^2+s^2-2\,r_\odot\,s\cos\psi}, \label{rspsi}
\end{equation}
$\psi$ is the angle between the direction in the sky and the
Galactic Center, $r_\odot\approx 8.0\,$kpc is the solar distance
from the Galactic Center, and $s$ the distance from the Sun along
the line-of-sight (l.o.s.). Particle physics enters via the term \be
f_{\gamma}(E)\,\Pi\equiv f_{\gamma}(E)\,\frac{\sigv}{2\,m_\chi^2},
\ee where $m_\chi$ is the mass of the WIMP (Weakly Interacting
Massive Particle), $\sigv$ is the annihilation cross section, and
the factor 1/2 enters since we assume that dark matter constitutes
its own anti-particle (if this is not the case, it should be
replaced by 1/4). The function $f_{\gamma}(E)$ is the photon
differential energy spectrum per annihilation (with units of
$E^{-1}$). This spectrum per annihilation is largely the result of
the fragmentation and hadronization of the WIMP's annihilation
products.

A general class of smooth halo distributions can be fitted as
\begin{equation}
\rho_{\rm sm}(r)=\rho_\odot\left(\frac{r_\odot}{r}\right)^\gamma
\left(\frac{r_\odot^\alpha+a^\alpha}{r^\alpha+a^\alpha}\right)^\epsilon,
\end{equation}
where $\rho_\odot$ is the dark matter density at the solar distance
from the Galactic Center, and $a$ is a characteristic scale radius
below which the profile scales as $r^{-\gamma}$ . Two of the most
well known profiles have been proposed by Navarro, Frenk and White
(NFW)~\cite{NFW}, with $\gamma=1\,,\alpha=1\,,\epsilon= 2$ and Moore
{\it et al.}~\cite{Moore} with
$\gamma=3/2\,,\alpha=1\,,\epsilon=3/2$. A cored isothermal profile
is recovered in the case of
$\gamma=0\,,\alpha=2\,,\epsilon=1$~\cite{isoth}.

These profiles differ primarily in the central region. Since here we
are focusing on the Galactic diffuse emission rather than that from
the Galactic Center, our choice of halo profile is not critical. The
uncertainties which are introduced through the choice of profile
(within a factor $\sim 2$) are negligible for our discussion. For
definiteness, we shall adopt an NFW profile with $\rho_{\rm
sm}(r_\odot)=0.3\,$ GeV/cm$^3$ and $a= 45\,$kpc.

We now turn to the sub-halos contribution to the average flux.
Unfortunately many details enter the calculation, including the
overall fraction of the dark matter mass which is still in the
clumps, the dark matter profile of the sub-halos, their mass
distribution, etc. However, it has already been shown in the past
(see e.g. L.~Bergstrom et al. in Ref. \cite{sub}) that their average
gamma-ray flux is mainly dependent on a single, effective parameter
(note that this is {\it not} the case for the statistical properties
of their angular fluctuations). To illustrate this point, let us
write the annihilation rate from a single sub-halo, $\Gamma_{\rm
cl}$, in terms of the density profile of the clump, $\varrho_{\rm
cl}$, as
\begin{equation}
\Gamma_{\rm cl}=\int \d^3 {\mathbf
x}\frac{\sigv}{2\,m_\chi^2}\varrho_{\rm cl}^2({\mathbf x})\equiv
\Pi\,M\,\varrho_0. \label{I1cl}
\end{equation}
The symbolic evaluation of the integral at the RHS has been written in terms of
a mass scale associated to the clump, $M$, and a density parameter,
$\varrho_0$. In general terms, the flux from the clumpy
component can be written as
\begin{equation}
I_{\rm cl}(E,\psi)=\frac{f_{\gamma}(E)}{4\pi}\int_{\rm l.o.s.} \d
s\int \d M \int \d p\: n_{\rm cl}[r(s,\psi),M;\,p]\,\Gamma_{\rm
cl}(M,p), \label{Icl}
\end{equation}
where $M$ is the mass of the sub-halos, $p$ a symbolic variable for
all the other unaccounted independent parameters, and
 $n_{\rm
cl}[r(s,\psi),M;\, p]$ is the number density distribution
(differential with respect to the $M$ and $p$) of sub-halos as a
function of $M,\,p$ and the distance from the Galactic Center, $r$.
The exact range over which the $M$ and $p$ integrals are carried out
depends indirectly on the cosmology and the particle physics model
which is considered. Equation (\ref{I1cl}) motivates one to write
\begin{equation}
\int \d M \int \d p\: n_{\rm cl}[r(s,\psi),M,p]\,\Gamma_{\rm
cl}(M,p)\equiv \varrho_{0}\,\zeta[r(s,\psi)]\, \rho_{\rm
sm}[r(s,\psi)]\,,\label{linear}
\end{equation}
where now $\varrho_{0}$ is purely a convenient normalization factor,
which we fix similarly to Ref.~\cite{Berezinsky:2003vn} as
$\varrho_{0}\equiv 10^{-21}\,$g cm$^{-3}$. For a constant $\zeta$,
the RHS of Eq.~(\ref{linear}) is the mathematical expression of the
hypothesis that the distribution of dark matter confined in
sub-halos traces the smooth distribution of the halo. In this case,
$\zeta$ can be thought as measuring the fraction of dark matter mass
which is still in the clumps. While $\zeta\sim const.$ is
approximately true at intermediate galactocentric distances, both
semi-analytic arguments and simulations suggests that $\zeta(r)\,
\rho_{\rm sm}(r)$ may tend to a cored distribution at low $r$, and
possibly decline at large $r$ as $r^{-2}$ instead of the $r^{-3}$
law expected both from the NFW and the Moore profile
\cite{Zentner:2003yd,Kravtsov:2004cm,Diemand:2004kx}. Equivalently,
one may have approximately that $\zeta(r)\propto r$ at large $r$ and
low $r$, while $\zeta \sim const.$ at intermediate galactocentric
distances. For simplicity we shall consider the $\zeta= const.$ as
our baseline case, but when relevant we shall comment on the
implications of the violation of this Ansatz in the following. Also,
the distribution of surviving sub-halos is especially uncertain at
small galactic radii, given the important role that tidal
interactions with baryons might play~\cite{tidal}. Since this is
still an open issue, to limit the impact of these uncertainties, we
conservatively discard the predictions of annihilation rates in the
inner regions $r\alt$3 kpc ($|\psi|\alt 20^\circ$). The residual
effect of interactions with disk stars will only affect the overall
annihilation rate by a factor of $\cal O$(1), which is subleading
with respect to other unknowns.

Under the previous assumptions, we can rewrite Eq.~(\ref{Icl}) as
\begin{equation}
I_{\rm cl}(E,\psi)=f_{\gamma}(E)\,\Pi\,\zeta\,\varrho_0\int_{\rm
l.o.s.} \d s\,\frac{\,\rho_{\rm sm}[r(s,\psi)]}{4\pi}. \label{Ic2}
\end{equation}

Combining the smooth and clumpy components, the total galactic signal can be written as
\begin{equation}
I_{\rm gal}(E,\psi)\equiv [I_{\rm sm}+I_{\rm
cl}](E,\psi)=\frac{1}{4\pi}f_{\gamma}(E)\,\Pi\int_{\rm l.o.s.} \d
s\,\rho_{\rm sm}[r(s,\psi)]\times\big(\rho_{\rm
sm}[r(s,\psi)]+\,\zeta\,\varrho_0 \big)\,.\label{Ic3}
\end{equation}
A numerical evaluation shows that even small degrees of substructure
(as low as $\zeta \sim 10^{-3}$) are sufficient to cause the
sub-halo contribution to dominate over the smooth component
everywhere but in the innermost region of the Galaxy. It is worth
commenting that this conclusion does not only hold for our Galaxy,
but also for other promising targets for dark matter annihilation
signals, like dwarf spheroidal (dSph) satellite galaxies of the
Milky Way. Although the geometry of the problem is different---in
the Galactic case we are {\it inside} the halo--- it can be shown
that the clumpiness in dSphs may be responsible for a global
enhancement with respect to the smooth halo flux of up to a factor
$\sim$100 (see Strigari et al. in Ref. \cite{gc}). For comparison, a
value of $\zeta\simeq 0.03$ would cause a similar enhancement for
the sub-halo to halo emission of our Galaxy in the direction
perpendicular to the Galactic Plane. Notice that, being $\varrho_0$
a convenient normalization constant chosen by hand, we have shown
that within simplifying but reasonable hypotheses the clumpy
contribution only depends on the single effective parameter $\zeta$.

On the other hand, the extragalactic contribution to the diffuse
gamma ray spectrum is given by~\cite{Bergstrom:2001jj}
\begin{equation}
I_{\rm ex}(E)= \frac{c}{4\pi}\Pi\int_0^\infty \d
z\,\frac{\rho_{\rm dm}^2(z)}{H(z)\,(1+z)^{3}}\,f_{\gamma}[E(1+z)]
\,e^{-\tau(E,z)}\,,\label{smoothedmap}
\end{equation}
where the Hubble function, $H(z)$, is related to the present Hubble
expansion rate, $H_0$, through the matter ($\Omega_M$) and the
cosmological constant energy density ($\Omega_\Lambda$), $H(z)=
H_0\sqrt{\Omega_M(1+z)^3+\Omega_\Lambda}$. The factor
$e^{-\tau(E,z)}$ accounts for the finite optical depth, $\tau$, of
the universe to high energy gamma rays due to scattering with the
extragalactic background light. The clumpiness of the dark matter is
usually taken into account by writing
\begin{equation}
\rho_{\rm dm}^2(z)\equiv\left(\frac{\rho_{\rm
dm,0}}{\rho_{c,0}}\right)^2\rho_{c,0}^2(1+z)^6
\Delta^2(z)=\Omega_{\rm dm}^2 \left(\frac{3 H_0^2}{8\pi G_N}\right)^2(1+z)^6
\Delta^2(z),
\end{equation}
where the so-called enhancement factor, $\Delta^2(z)$, can be parameterized approximately as
(see, for example, Ref.~\cite{Bergstrom:2001jj})
\begin{equation}
\Delta^2(z)=\frac{\Delta^2(0)}{(1+z)^3},
\end{equation}
where $\Delta^2(0)$, at least for $z\alt 10$, has only a weak
dependence from $z$. This is confirmed in more detailed
calculations, as the one in Ullio et al. in Ref. \cite{gc}. Their
Fig. 5 clearly shows that $(1+z)^3\,\Delta^2(z)/H(z)\propto 1/H(z)$,
apart for the range $z\alt 1$ where a residual dependence of of
${\cal O}$(1) may be present, depending on the model considered.
Since this is a sub-leading uncertainty for our level of
approximation, we will neglect any $z$ dependence in $\Delta^2(0)$.
According to semi-analytical estimates, $\Delta^2(0)$ can have
values ranging from $10^4$ to $10^8$~\cite{Taylor:2002zd}. All
together,
\begin{equation}
I_{\rm ex}(E)= \frac{c}{4\pi}\,\Pi\,\Omega_{\rm dm}^2\,\rho_{c,0}^2\,\Delta^2(0)\int_0^\infty \d
z\,\frac{f_{\gamma}[E(1+z)]\,e^{-\tau(E,z)}}{H_0\sqrt{\Omega_M(1+z)^3+\Omega_\Lambda}}\,
\,.\label{smoothedmap2}
\end{equation}

\begin{figure}[!tbp]
\epsfig{file=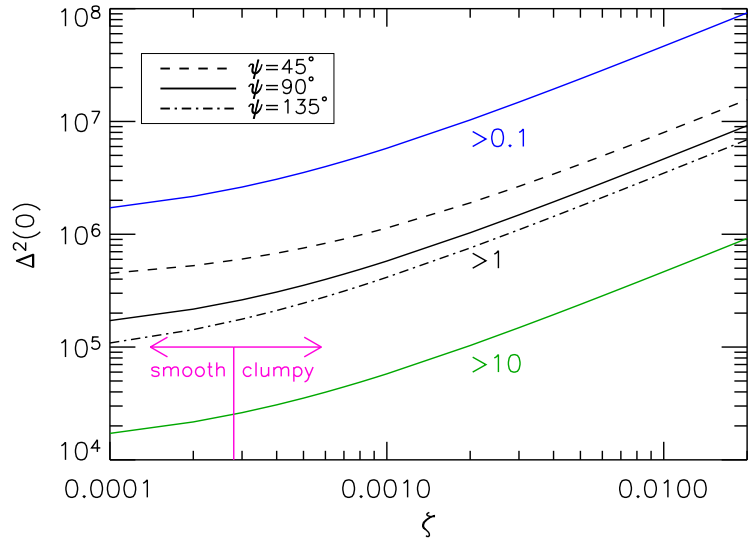,width=.7\columnwidth} \caption{The contours
$F_{\rm gal}/F_{\rm ex}=0.1,1,10$ in the direction orthogonal to the
Galactic plane ($\psi=\pi/2$) in the $\zeta-\Delta^2(0)$ plane, as
well as the contour $F_{\rm gal}/F_{\rm ex}=1$ for three values of
$\psi$. In the lower-right part of the plot, the Galactic radiation
dominates the dark matter contribution to the diffuse flux. To the
right of the purple vertical line, the contribution from sub-halos
dominates over the smooth halo term at $\psi=\pi/2$.}
\label{Contour}
\end{figure}

The relative weight of the Galactic and extragalactic contributions
has only a weak dependence on the spectral shape, $f_\gamma$. In the
following, we shall consider the integral quantity, $F_{\rm
i}(\psi)\equiv \Pi^{-1}\int\d E I_{\rm i}(\psi)$. We also adopt, for
definiteness, $m_\chi=100\,$GeV whenever not specified otherwise.

In Fig. \ref{Contour}, we show in the $\zeta-\Delta^2(0)$ plane the
three contours $F_{\rm gal}/F_{\rm ex}=0.1,1,10$ in the direction
orthogonal to the Galactic plane ($\psi=\pi/2$), where the
astrophysical contribution to the diffuse flux is minimized. For the
$F_{\rm gal}/F_{\rm ex}=1$ case, we also show results for two other
values of $\psi$ ($\pi/4$ and $3\pi/4$). In the lower-right region
of the plot, the Galactic radiation dominates the dark matter
annihilation contribution to the diffuse flux (the absolute flux
depends on the factor $\Pi$). To the right of the purple vertical
line, the contribution from sub-halos dominates over the smooth halo
term at $\psi=\pi/2$.

Although very rough, the simple model sketched above is useful for
understanding qualitatively some features, as illustrated by the
following example. In Ref.~\cite{Elsaesser:2004ap}, the claim is
made that a dominant fraction of the high energy diffuse gamma ray
spectrum measured by EGRET~\cite{Sreekumar:1997un,Strong:2004ry} is
consistent with being due to $m_\chi\sim 500\,$GeV annihilating
neutralinos, without violating existing constraints.
Ref.~\cite{Ando:2005hr}, in contrast, concludes that it is hardly
conceivable that dark matter annihilation could be a main
constituent of the extragalactic background without exceeding the
observed gamma ray flux from the Galactic Center, at least if the
density profile of the Milky Way is not very different from that
found in other galaxies. We note here that the two statements are
equally valid within the presently allowed parameter space, although
they assume very different priors. The enhancement factor considered
in Ref.~\cite{Elsaesser:2004ap} is around $\Delta^2(0)\sim 10^7$
(where the extragalactic flux is likely to dominate for any
reasonable value of $\zeta$), while in Ref.~\cite{Ando:2005hr},
values of $\Delta^2(0)\sim 10^4-10^5$ are assumed (for which the
galactic flux dominates over the extragalactic one for any
reasonable value of $\zeta$). Although the sub-halo profiles assumed
are the same, in \cite{Ando:2005hr} it was noted that such a large
discrepancy may come from uncertainties concerning the concentration
parameter and the presence/absence of substructures. Whatever the
cause, note that already a boost of $10^2-10^3$ in the enhancement
factor would bring the expectation of Ref.~\cite{Ando:2005hr} for
the extragalactic contribution within an order of magnitude of the
EGRET flux, similar to the findings of Ref.~\cite{Elsaesser:2004ap}.
Although, as correctly noted in Ref.~\cite{Ando:2005hr}, the
parameters that we call here $\zeta$ and $\Delta^2(0)$ are to some
extent correlated, it is plausible that a ``large" $\zeta$ value
might not affect significantly the Galactic Center signal, but only
the Galactic diffuse signal, due to the effects of tidal disruption,
for example. Namely, the Galactic Center flux, which has the highest
uncertainty due to the unknown extrapolation of the smooth halo
profile, has probably no significant contribution from the clumpy
fraction of the halo.

\section{Distinctive Patterns of a dominantly extragalactic contribution}\label{extragal}
In this section, we discuss the most prominent features in the
angular distribution of the diffuse gamma rays if the extragalactic
contribution dominates. Although the extragalactic dark matter flux
should carry interesting statistical information (see
\cite{Ando:2005xg,Ando:2006cr}; also, see
\cite{Ando:2006mt,Miniati:2007ke} for a discussion of astrophysical
backgrounds) we shall concentrate our analysis here on {\it
deterministic} features, especially at large scales.

\subsection{Cosmological Compton-Getting Effect}\label{CCG}
An observer in motion with velocity, $\bu$, relative to the
coordinate system in which the distribution of gamma rays is
isotropic will measure an anisotropic flux. If gamma ray sources
are, on average, at rest with respect to the cosmological frame, the
magnitude and direction of the velocity, $\bu$, of the solar system
can be deduced from the detection of the dipole anisotropy of the
CMB,  $u=369\pm 2$~km/s in the direction
$(l,b)=(263.86^\circ,48.24^\circ)$~\cite{Yao:2006px}. Since
$u\equiv|\bu|\ll 1$, the anisotropy is dominated by the lowest
moment, {\it i.e.} its dipole moment. The Galactic analogue of this
effect was first noticed as a diagnostic tool for cosmic rays by
Compton and Getting~\cite{CGpaper}.

The amplitude of this anisotropy can be derived from the Lorentz
invariance of the the phase space distribution function, $f$, in the
frame of the observer and of the emitters (see, for example,
Ref.~\cite{Kachelriess:2006aq}). A first order expansion allows one
to deduce the shape of the differential intensity. If we denote with
$E$ the energy in the emission (rest) frame, and with $\Ep$ the
energy as measured in the observer (moving) frame, the intensity in
the moving frame can be written as
\begin{equation}
\Ip(E,{\mathbf n})\simeq I(E)\left[1+\left(2-\frac{\d\ln I}{\d\ln
E}\right)\bu \cdot {\mathbf n}\right]\,,
\end{equation}
where in the above formula ${\mathbf n}$ is the generic direction and
the identification of the energy in the two frames $E=\Ep$ has been made
consistently with the first order result. The function, $I(E)$, is the number of
particles per unit solid angle and unit energy that pass per unit of
time through an area perpendicular to the direction of observation, and
it is related to the phase space density, $f$, by $I(E)\simeq E^2f(E)$.
Thus a dipole anisotropy is expected with an
amplitude given by
\begin{equation}
A\equiv \frac{I_{\rm max}-I_{\rm min}}{I_{\rm max}+I_{\rm
min}}=\left(2-\frac{\d\ln I}{\d\ln E}\right)\,u
\,,\label{Ianisamplitude}
\end{equation}
which is independent of energy as long as the energy spectrum does
not change, {\it i.e}. when it can be approximated by a single
power-law. Taking into account the observed spectrum of diffuse
gamma rays $I(E)\propto E^{-2.1}$~\cite{Sreekumar:1997un}, one
infers $A=(2+2.1)\,u\simeq 0.5\%$. Note that the Earth's motion with
respect to the Sun induces a subleading (8\%) modulation in the
vector $\bu$.

Unfortunately, this is not characteristic of a dark
matter signal, but would be shared by any cosmic distribution of
sources. Yet, its detection (or limits on its amplitude) would allow
one to constrain the fraction of the diffuse gamma ray background
emitted by cosmological sources~\cite{Kachelriess:2006aq}, which
would translate into a conservative constraint on the product,
$f_\gamma\times\Pi \times \Delta^2(0)$.

\subsection{Intrinsic Anisotropies}
The distribution of the dark matter in the present day universe is
highly structured. As a result, intrinsic anisotropies should be
present in the diffuse gamma ray background from dark matter
annihilations. Recently, it has been proposed that one could use the
peculiar anisotropy at small-scales to probe dark
matter~\cite{Ando:2005xg}. Provided that dark matter contributes in
a relevant way to the diffuse flux, this signature has promising
chances to be detected by GLAST, the Gamma Ray Large Area Space
Telescope to be launched at the end of this year by NASA. This
interesting conclusion was recently extended in
Ref.~\cite{Cuoco:2006tr}, which notes that most of the anisotropy
pattern at relatively large angular scales is due to nearby large
scale structures. This is especially true at energies higher than
${\cal O}$(0.1-1) TeV, when gammas start to be absorbed through pair
production on the extragalactic background light, and the observable
horizon shrinks considerably. This means that one can use the whole
information of a real large-scale structure catalog, going far
beyond mere statistical predictions. How strong the correlation
between gamma-ray sources and the large scale matter potential is
depends on the bias of the sources with respect to the catalogue
tracers. In general, it is present both for astrophysical and dark
matter models for the production of gamma rays, although the
correlation with overdensity should be stronger for dark matter
annihilation radiation. For high energy photons, pair production
energy loss is sufficiently effective such that the anisotropies
depend almost entirely on the structure of dark matter in the local
universe (within several hundred Mpc or so). In this case,
cosmological uncertainties, the energy spectrum of the dark matter
signal (in dark matter models), the redshift evolution of the
sources, etc., each play subleading roles. Additionally, while the
``local" anisotropic flux is almost unchanged, at increasingly high
energies, the far isotropic component is cut away more and more. The
relative anisotropy is higher the greater the energy cut. Together
with the many modes available ($2l+1$ for each multipole
coefficient, $C_l$) and the absence of the cosmic variance
limitation, this compensates, to some extent, for the lower
statistics available in the $E\agt 0.1\,$TeV range. Of course, the
strength of this technique depends critically on the mass of the
WIMP being considered. See Ref.~\cite{Cuoco:2006tr} for further
details.

\begin{figure}[!tbp]
\epsfig{file=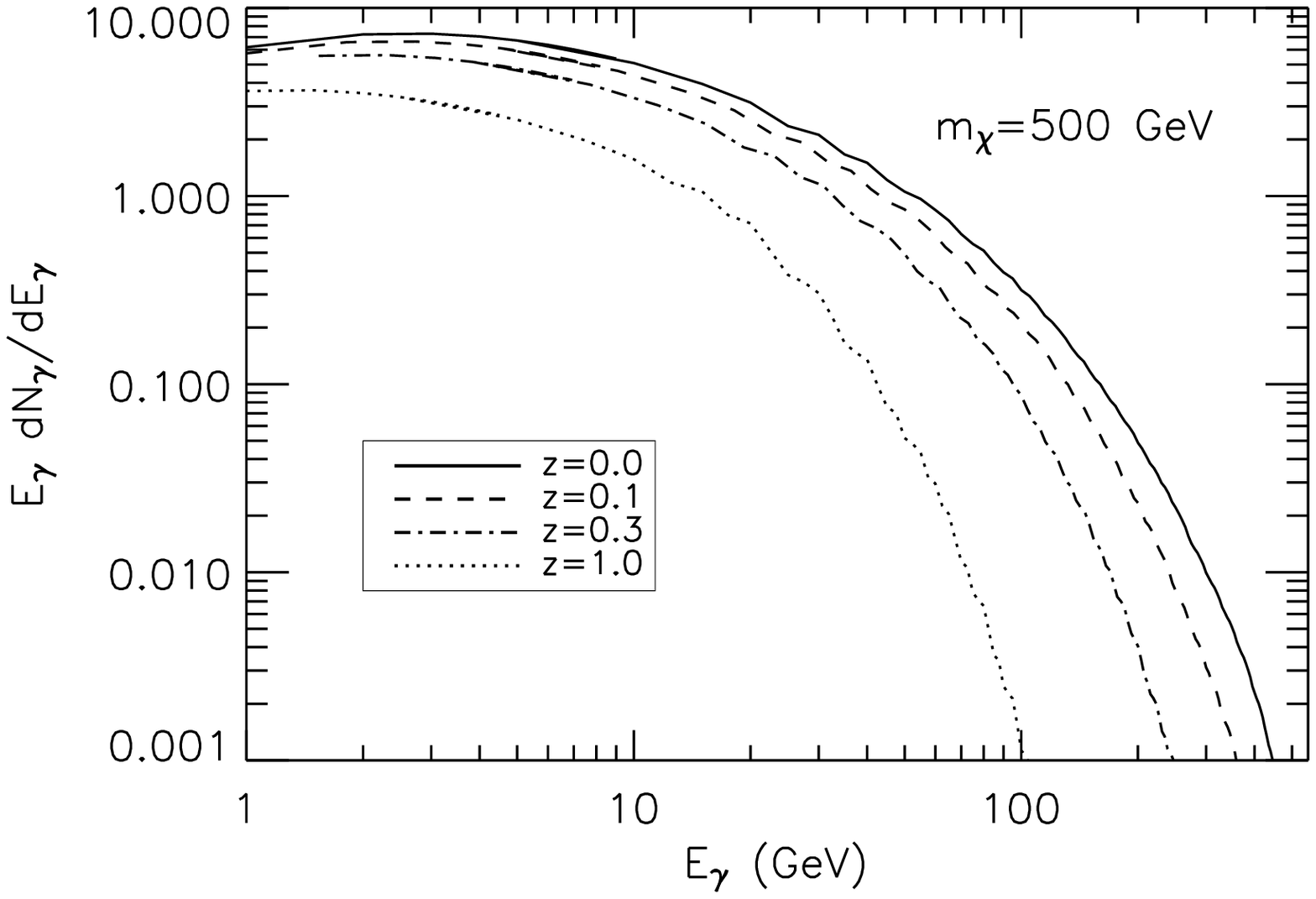,width=.49\columnwidth}
\epsfig{file=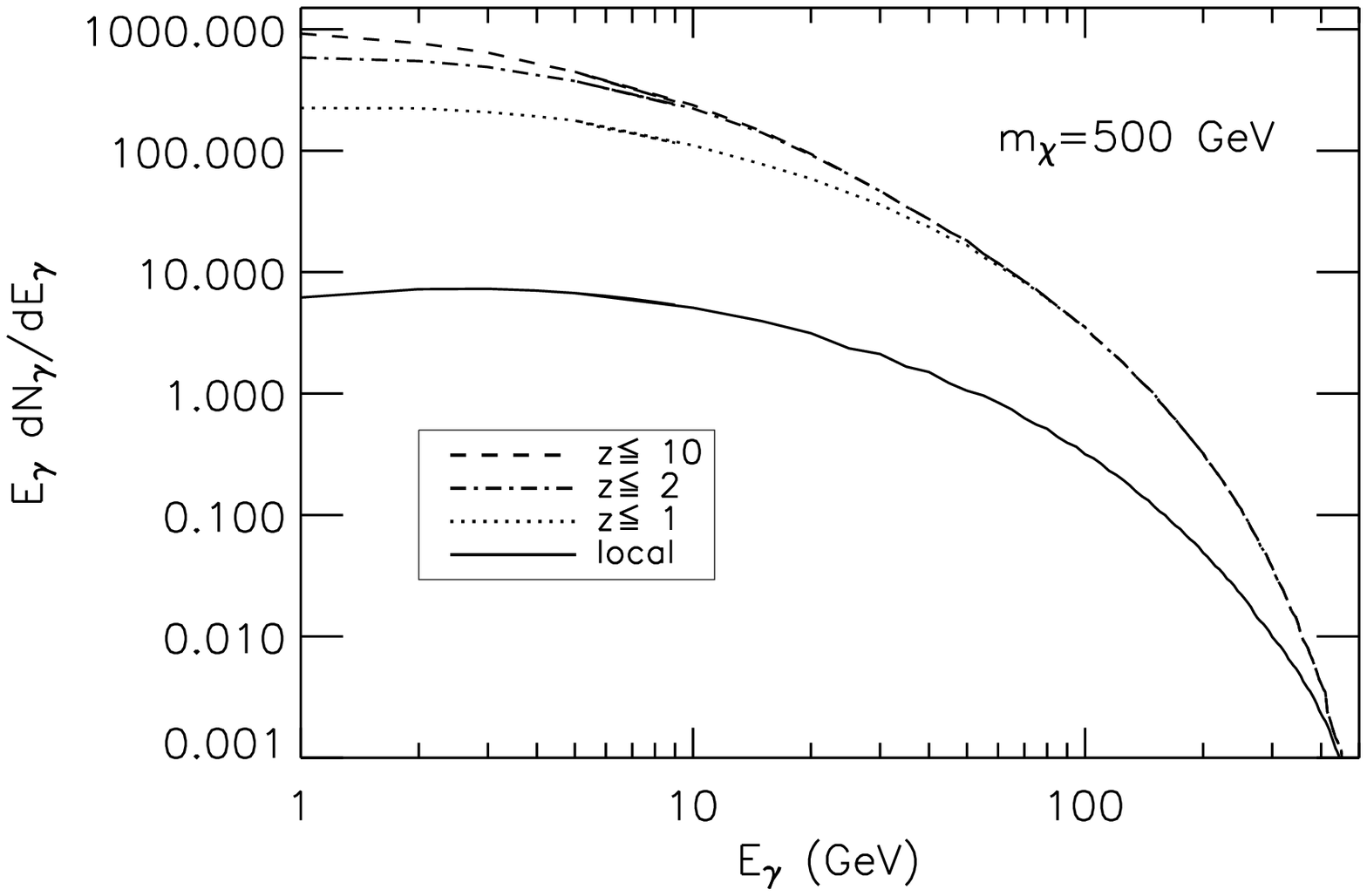,width=.49\columnwidth} \caption{Left: The
shape of the gamma ray spectrum from dark matter annihilations from
four redshifts (0, 0.1, 0.3 and 1.0), including the effects of
redshift energy losses and absorption with the cosmic infrared
background. In the right frame, we compare the spectral shape of
gamma rays from local dark matter annihilations (solid) to that from
sources integrated out to a redshift of 1, 2 and 10 (from
bottom-to-top, dashed). We have assumed that $\Delta^2 (0)$ is
roughly constant over this range of redshifts. In each frame, a 500
GeV WIMP annihilating to $W^+ W^-$ was considered. The cosmic
infrared background spectrum found in Ref.~\cite{ahrirb} was
adopted.} \label{irb}
\end{figure}

Additionally, the shrinking of the observable horizon for gamma rays
with energies higher than ${\cal O}$(0.1-1) TeV has important
consequences for the spectral shape of a dark matter signal if
$m_\chi$ is large enough. In the left frame of Fig.~\ref{irb} we
plot the spectral shape of gamma rays produced in dark matter
annihilations at four redshifts, $z=$0, 0.1, 0.3 and 1.0. Here, we
have considered a 500 GeV WIMP annihilating to $W^+ W^-$. In the
right frame, we show the shape of the gamma ray spectrum from dark
matter annihilations integrated out to a redshift of $z=$1, 2 and 10
(from bottom-to-top, dashed), compared to the harder spectral shape
emitted at redshift zero, e.g. from a local source (solid). Here, we
have assumed that $\Delta^2 (0)$ is constant in this range of
redshifts. Also, we have matched the integrated spectra to the local
one at $E_\gamma=m_\chi$, in order to show clearly how the redshift
distribution softens the continuum spectrum. In each frame, we have
adopted the model of the cosmic infrared background spectrum found
in Ref.~\cite{ahrirb}. The optical depth of the universe to high
energy gamma-rays is calculated by simply integrating over the cross
section for pair production and the spectrum of the infrared
background.

\section{Distinctive Patterns of a dominantly galactic contribution}\label{gal}
It is well known that the angular distribution of gamma radiation
from dark matter annihilation in the Galactic halo has potentially
interesting signatures (see, for example, the seminal paper,
Ref.~\cite{Calcaneo-Roldan:2000yt} or the discussion based on the
recent simulations reported in J.~Diemand et al. of Ref.
\cite{sub}). In the following, we summarize the basic properties
expected, once again emphasizing their deterministic features.

\subsection{Offset Position of the Sun}\label{offset}
\begin{figure}[!tbp]
\epsfig{file=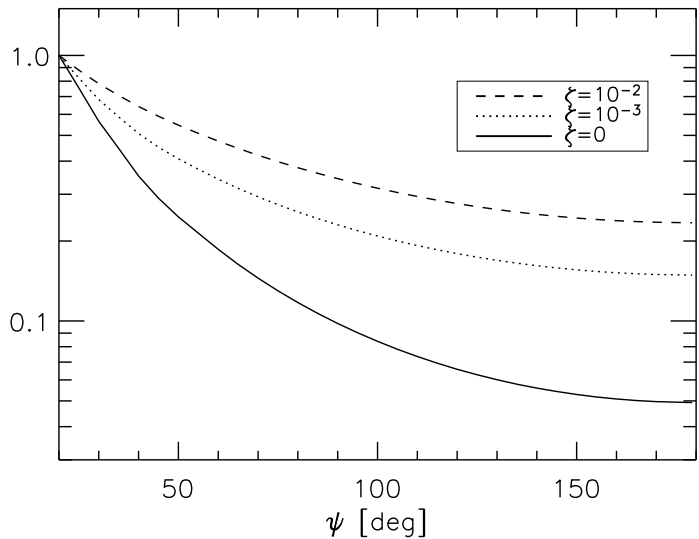,width=.7\columnwidth}
\caption{The angular profile of the Galactic emission of gamma's from
dark matter annihilation for three values of the clumpiness parameter, $\zeta$.
The curves have been normalized to 1 at $\psi=20^{\circ}$. The region $\psi<20^{\circ}$
is not shown, since it is affected by large uncertainties (see text for discussion).} \label{Profile}
\end{figure}

In the limit of exact spherical symmetry of the dark matter halo, an
observer placed in the center of the distribution would observe an
isotropic annihilation signal. This follows trivially from
Eq.~(\ref{rspsi}) in the limit $r_\odot\to 0$. However, the Sun is
offset with respect to the center of the Galactic halo. This induces
a peculiar angular dependence, with a maximum toward the inner
Galaxy and a minimum toward the antigalactic Center. We show this in
Fig.~\ref{Profile} for three values of $\zeta$. In the limit of
spherically symmetric halo, this is independent of the azimuthal
angle, $\phi$, around the Sun-Galactic Center line. This signature
is fairly robust in two limits:
\begin{itemize}
\item[A)]
When $\zeta \alt 0.0002$ (and $\Delta^2(0)\alt 10^5$)
and the angular distribution is dominated by the smooth halo (bottom curve
in Fig. \ref{Profile}). For example, the ratio of the flux at $\psi=45^\circ$
to the flux at $\psi=135^\circ$ is about 5.

\item[B)] When $\zeta \agt 0.0005$ (and $\Delta^2(0)\alt {\rm few}\times 10^5$)
and the angular distribution is dominated by the clumpy halo (top
curve in Fig \ref{Profile}). In the limit of a constant $\zeta$, the
ratio of the flux at $\psi=45^\circ$ to the flux at $\psi=135^\circ$
is of about 2.2.
\end{itemize}

Note that the two models considered in Ref.~\cite{Aloisio:2002yq}
correspond to these two extreme situations. It is however worth to
point out that the quantitative conclusions in the second case
($\zeta\agt 0.0005$) are not robust if one drops the assumption of a
quasi-constant $\zeta(r)$. In particular, a decreasing $\zeta(r)$
within the solar circle would partially suppress the anisotropy, and
paradoxically one may end with an almost isotropic flux despite the
offset position of the Sun. A quantitative estimate of such effect
is unfortunately model-dependent, in particular on the radial
dependence of the tidal-stripping effects of dark matter sub-halos,
including the non-negligible effects of the baryonic material in the
inner solar system. For details on this (still) open issue, we
address to Ref.~\cite{tidal}.

\subsection{Compton-Getting Effect (Proper Motion of the Sun in the Halo)}\label{haloCG}
A second signature, which to the best of our knowledge is discussed
in detail here for the first time in relation to dark matter, is
purely kinematic. Unlike the disk of our Galaxy, which is supported
against radial collapse by its angular momentum, the dark halo is
supported by random velocities which serves as a collisionless
pressure. It is expected that the velocities of dark matter
particles were isotropized at the time of the formation of the
Galaxy in the so-called process of ``violent relaxation"
\cite{Lynden-Bell:1966bi}, and should retain this distribution as
long as no relevant interaction with the collisional baryonic gas
intervenes (for a more extended discussion, see
Ref.~\cite{Drukier:1986tm}). Although the halo may have some
rotation, we know for sure that it is not supported by angular
momentum, otherwise it would be flattened similarly to the disk. For
our Galaxy, observational constraints imply that the halo is almost
spherical, with differences between the axes of the best-fit
spheroid not larger than $\sim 20$\%
\cite{Olling:1999ss,Olling:2001yt}. If we neglect an intrinsic
angular momentum of the halo, our motion with respect to the halo is
fully due to the Galactic disk rotation around the center of the
Galaxy with a velocity of about 220 km/s at the solar galactocentric
distance (see e.g. Ref.~\cite{Olling:2001yt} and references
therein). Analogously to the previously discussed extragalactic
case, this would be manifest as a halo Doppler effect. In this
Galactic case, the dipole points in the direction of motion of the
rotation of the Galaxy, and with an amplitude of about 0.3\% [see
Eq.~(\ref{Ianisamplitude})]. The amplitudes increase when the
spectrum softens, and may be very large (although in a narrow energy
range) close to an abrupt cutoff in the energy spectrum, as the one
for $E_\gamma\simeq m_\chi$. Note that in the case of a Galactic
astrophysical (non-dark matter) origin of the diffuse gamma ray
flux, this effect is not present, since the disk is on average
co-moving with the Sun, with typical proper velocities much smaller
that the bulk one. This signature is, therefore, specific to dark
matter, unlike its extragalactic counterpart described in
Sec.~\ref{CCG}.

\subsection{Asphericity of the Halo}
Thus far, we have implicitly assumed that the dark matter
distribution in the halo has a spherical symmetry. Simulations of
the halo structure, however, show some departure from the spherical
symmetry, and typical dark matter halos are actually spheroidals
with a moderate eccentricity. Also, the level of ellipticity may be
radially varying, with larger ellipticities towards the inner
Galaxy, see e.g. \cite{Hayashi:2006es}. Note however that there are
hints that the inclusion of baryonic dissipation may produce
significantly rounder halos than those formed in equivalent
dissipationless simulations especially in the inner halos
\cite{Kazantzidis:2005ru}.

The impact of this effect on the annihilation flux was
discussed in detail in Ref.~\cite{Calcaneo-Roldan:2000yt}, and we shall
not repeat here that analysis. Qualitatively, the anisotropy
in the emission may  be comparable to the effect
of the offset position of the Sun. However the symmetry axes of the halo
are unknown. Thus the magnitude, the exact angular dependence and
the direction of this anisotropy are unknown, making identification of this effect challenging.

\subsection{Proper Anisotropy Due to Clumpiness}
As in the case of extragalactic emission, the typical parameters of
the clumps are reflected as peculiar angular signatures of the
annihilation signal in the sky. In contrast to the signatures
considered above, this is most important at small scales, unless a
few clumps happen to be very near us (and in the case that they are
resolved, it may also be possible to detect their proper
motion~\cite{Koushiappas:2006qq}). While a highly clumpy and steady
emission may be a spectacular signature of dark matter annihilation,
it is difficult to obtain model-independent predictions on the
expected angular power spectrum, since the physical processes
determining the result are highly non-linear and entangled (such as
mergers, tidal disruptions, etc.). Furthermore, the observer-related
``galactic variance"---the dependence of the signal features on the
actual distribution of clumps around our position, which is
unknown---may greatly change the expected signal. Some attention has
previously been given to this kind of signature within the context
of superheavy dark matter decays connected with the ultra-high
energy cosmic rays~\cite{Blasi:2000ud}.

\section{Prospects for GLAST}\label{detection}
Regarding the detectability of dark matter annihilations through the
various angular signatures described in this paper, we shall limit
our discussion to the GLAST satellite detector~\cite{glast}.
Although the current generation of atmospheric Cerenkov telescopes,
including H.E.S.S.~\cite{hess}, MAGIC~\cite{magic} and
VERITAS~\cite{veritas}, are highly sensitive to very high energy
gamma rays, possibly including dark matter radiation, they are best
suited for studying point sources rather than diffuse emission.

\begin{figure}[!tbp]
\epsfig{file=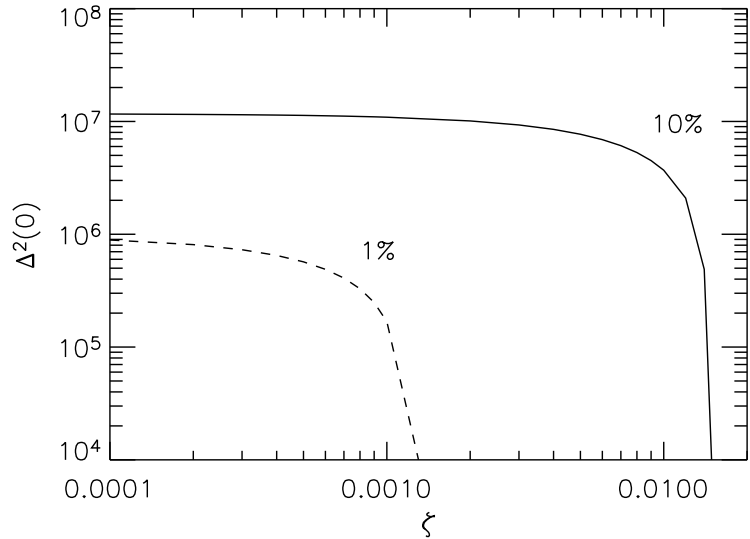,width=.7\columnwidth} \caption{Contours in
the $\zeta-\Delta^2(0)$ plane (see Sec.~\ref{galexgal}) illustrating
the fraction of the diffuse flux observed by EGRET (above 1 GeV)
which is the product of dark matter annihilations. Above and to the
right of the solid line, more than 10\% of this flux is produced by
dark matter. Above and to the right of the dashed line, dark matter
generates 1\% or more of this flux. Here, we have used a 100 GeV
WIMP with a $3 \times 10^{-26}$ cm$^3$/s annihilation cross section.
The Galactic flux being considered here is at a direction
perpendicular to the Galactic plane.} \label{Contour2}
\end{figure}

The flux of gamma rays due to dark matter annihilation is at present
unknown. To the best of our knowledge, it may account for a
significant or even dominant fraction of the diffuse gamma ray
background, especially at the highest measured
energies~\cite{Elsaesser:2004ap}. A residual, isotropic radiation
usually interpreted as extragalactic gamma ray background has been
measured by GLAST's predecessor, EGRET~\cite{Sreekumar:1997un}. The
intensity of this spectrum can be fit by~\cite{Sreekumar:1997un} (see
also Ref.~\cite{Strong:2004ry})
\begin{equation}
I_{\rm cosmic}(E_{\gamma})=(7.32\pm 0.34)\times 10^{-6}
\left(\frac{E_{\gamma}}{0.451 {\rm GeV}}\right) ^{-2.10\pm0.03} {\rm
cm}^{-2}{\rm s}^{-1}{\rm sr}^{-1}{\rm GeV}^{-1},
\label{spectrum98}
\end{equation}
over an energy range from $E_{\gamma}\sim\,$10 MeV to $E_{\gamma}\sim\,$100 GeV.
The number of events collected above an energy,
$E_\gamma$, can be written as
\begin{equation}
N_\gamma = t\cdot\Omega_{\rm fov}\cdot
\int_{E_\gamma}^\infty{\rm d}E\, A_{\rm eff}(E)I_{\rm cosmic}(E)
\,, \label{Ngamma}
\end{equation}
where $\Omega_{\rm fov}$ is the solid angle of the field-of-view,
$A_{\rm eff}(E)$ is the effective collecting area of the instrument
(averaged over its field-of-view), and $t$ is the time observed. The
LAT detector on board GLAST features $A_{\rm eff}(E)\simeq
10^{4}\,$cm$^2$ for $E\agt 1\,$GeV (and $A_{\rm
eff}(E)>4000\,$cm$^2$ for $E\agt 0.1\,$GeV), $\Omega_{\rm fov}\simeq
2.4\,$sr, and excellent hadronic background rejection capabilities
\cite{glast}. Even when considering realistic fractions of data to
be rejected from the use of mask cuts or a non-ideal duty-cycle, at
least $\cal{O}$(10$^6$) diffuse photons per year should be
measurable in the GeV range, given the background flux measured by
EGRET. Considering up to a decade of operation, it is clear that as
long as dark matter is responsible for a significant fraction of the
flux of Eq.~(\ref{spectrum98}), say at least 10-15\%, GLAST should
be sensitive to anisotropies down to the few permil level. Modulo
the issue of foreground removal, this precision should be sufficient
to detect the tiniest of the signatures considered in this paper.
For illustrative purposes, in Fig.~\ref{Contour2} we show the
regions in the parameter space where the DM flux amounts to more
than 10\% or 1\% of the EGRET flux given by Eq.~(\ref{spectrum98}),
assuming typical parameters for the DM candidate, and fluxes
integrated above 1 GeV. In the region above the solid line the
collected statistics would allow one to detect all the signatures
discussed in this article. Most of these signatures should be
detectable in the region above the dashed line. Interestingly, in
the lower left corner of the parameter space where the DM flux is
subdominant, the signatures of the smooth halo signal should be
prominent. Since the offset position of the Sun implies anisotropies
of ${\cal O}$(100\%), even in this range the prospects for the
detection of DM signatures are more promising than what one would
naively deduce from Fig.~\ref{Contour2}.

Of course, a major obstacle in detecting such features are
astrophysical foregrounds, which can make the identification of the
above signatures difficult (the same is true, of course, when
considering spectral energy features). The very subtraction of the
Galactic astrophysical foreground is a highly non-trivial issue, see
e.g. \cite{Strong:2004ry}. For the purpose of detection of the
diffuse dark matter flux, apart for applying low-latitude cuts (the
astrophysical contamination is higher at low galactic latitudes), it
may be appropriate to fit simultaneously for the sum of Galactic
astrophysical flux (as provided by numerical routines like GALPROP
\cite{Galprop}) plus an additional dark matter component. In
addition, there are fortunately a number of qualitative properties
of dark matter signals which may help in distinguishing them from
astrophysical emissions. The strategy to reveal peculiar
extragalactic features has been discussed
elsewhere~\cite{Ando:2006cr,Cuoco:2006tr} and we shall not repeat it
here. We want only to stress that, qualitatively, after bright
sources are removed, a significant cross-correlation with large
scale structure catalogs or the CMB dipole direction would provide
important diagnostics and a tool for background rejection.

In case of dominance of the Galactic dark matter halo emission, it is important
to note that: (i) The offset position of the Sun produces a
$\psi$-dependent asymmetry. (ii) The Doppler effect due to our
motion in the halo produces a dipolar asymmetry in the azimuthal
angle, $\phi$, which for a fixed $\psi$ identifies the direction
around the Sun-Galactic Center axis (each corona should share the
small dipole in the direction of the motion of the disk in the
halo). Note also that, although small, its angular shape and
direction are known a priori, which facilitates a search for it.
(iii) The signature of the asphericity of the halo depends on the
unknown shape of the halo. On general grounds, one should expect a
modulation of the emission in the angle, $\psi$, (as does the one
treated in Sec.~\ref{offset}), as well as a modulation in the
azimuthal angle, $\phi$ (as does the one treated in
Sec.~\ref{haloCG} which, however, should be smaller). Barring
fine-tuned directions of the halo axes of symmetry, a more specific
signature is an asymmetry pattern between the $\psi,\phi$ and
$-\psi,\phi$ directions (reflection with respect to the Galactic
plane), which cannot be mimicked by any of the two previously
considered effects. Note that, although the relative magnitude of
the anisotropy measured may change (whenever dark matter does not
constitute a constant fraction to the diffuse gamma signal), the
relative weights of the different Galactic asymmetries discussed
here remain constant with energy. Since the diffuse radiation will
be observed in many energy bands, one should keep in mind that the
shape of this pattern must stay constant at each energy if it is to
be attributed to dark matter. Also, if suspected dark matter
spectral features appear, they should correlate with the amplitude
of the angular asymmetries discussed in this paper. Combining
angular and energy spectral information should surely strengthen our
ability to distinguish dark matter annihilation radiation from other
types of sources. Indeed, there is no a priori reason to expect that
an astrophysical foreground affecting either the energy spectrum or
the angular spectrum should appear in the other as well.

\section{Conclusions}
\label{summary}

With the next generation satellite-based gamma ray telescope, GLAST,
our measurements of the diffuse gamma ray spectrum will improve
dramatically. Not only will the number of events observed increase,
but also the spectrum will be studied up to considerably higher
energies. Furthermore, much of the astrophysical contributions to
the diffuse flux observed by EGRET will likely be resolved as point
sources as a result of GLAST's superior angular resolution.

In this article, we have studied the signatures and distinctive
features that would be possessed by dark matter annihilation
radiation from either a dominantly Galactic or extragalactic
population of dark matter. For an extragalactic contribution, we
discussed the effect of observing dark matter radiation in a moving
frame of reference (relative to the dark matter distribution), known
as the Compton-Getting effect. We also discuss the ability to tie
the angular distribution of annihilation radiation to the known
structure of our (cosmologically speaking) local universe. This is
especially interesting for gamma rays more energetic than $\sim$100
GeV, which can be absorbed by the cosmic infrared background
radiation, effectively reducing the horizon for such particles.

If the diffuse dark matter annihilation radiation is instead
dominated by the Galactic population, this flux will contain
distinctive features resulting from the offset location and proper
motion of the Solar System relative to the Galactic halo.
Furthermore, any asphericity in the halo will have effects of the
angular distribution of annihilation radiation. As in the
extragalactic case, we also briefly discuss the anisotropies which
are likely to result from inhomogeneities and substructure in the
Galactic dark matter distribution.

GLAST's predecessor, EGRET, observed the presence of a diffuse gamma
ray spectrum. Based on these observations, GLAST should see
$O(10^6)$ diffuse photons per year above the GeV \footnote{At high
latitudes and for energies $E\agt\,$GeV, the diffuse EGRET flux is
comparable or larger than the Galactic flux reported in
\cite{Hunter:1997}, as it can be easily checked with the fit
provided by  L.~Bergstrom et al. in Ref \cite{gc}.}. If a
significant fraction of this flux is the product of dark matter,
many of the angular features described here could potentially be
identified. For example, if 10-15\% of the diffuse flux observed by
EGRET is dark matter annihilation radiation, in a decade of
operations GLAST should be sensitive to anisotropies down to the few
permil level, comparable or often better than the level required to
study the angular signatures we have discussed in this paper.

\bigskip

\section*{Acknowledgments}
This work has been supported by the US Department of Energy
and by NASA grant NAG5-10842.

\appendix
\section{Flux Enhancement due to clumpiness}\label{simpleargument}
Consider a cubic volume of size $L$ at a distance $D\gg L$ from us,
such that the solid angle subtended by its surface is $\Theta^2$,
where $\Theta\equiv L/D$. Let us assume that it is filled with: (i)
a homogeneous distribution of dark matter with density $\bar\rho$;
(ii) a cubic dark matter clump of size $l\ll L$ and density
$\rho_c$, having the same mass ({\it i.e.}, such that $\rho_c\,
l^3=\bar\rho\,L^3$). We denote with $\bar I$ and $I_c$ the
differential photon fluxes (per area, per time, per solid angle)
collected in the two cases, and with $\bar J=\bar I\,\Theta^2$ and
$J_c=I_c\,\theta^2$ the flux per area per time, where $\theta=l/D$.
If the angular resolution of our instrument is larger than $\Theta$,
then the quantities $\bar{J}$ and $J_c$ are the relevant
observables.

Let us assume now that $I\propto \int_{\rm l.o.s.}\rho^{\kappa}\d
x$, where $\kappa=1$ corresponds to a decaying dark matter scenario
and $\kappa=2$ to an annihilating dark matter one. One immediately
sees that
\be
I_c\propto\rho_c^{\kappa}\,l=\bar \rho^{\kappa} \left(\frac{L}{l}\right)^{3\kappa}l
=\bar \rho^{\kappa} \left(\frac{L}{l}\right)^{3\kappa-1}L\,,
\ee
and
\be
I_c=\bar I  \left(\frac{L}{l}\right)^{3\kappa-1}\,,
\ee
from which it follows
\be J_c=\left(\frac{L}{l}\right)^{3\kappa-3}\bar
J=\left(\frac{\rho_c}{\bar\rho}\right)^{\kappa-1}\bar
J\,.\label{eqa3}\ee
Equation (\ref{eqa3}) implies that when $\kappa=1$ (the decaying
dark matter scenario), only the angular information is sensitive to
the clumpiness of the distribution. In other words, the total photon
flux from decaying dark matter can be calculated (for a fixed
particle physics scenario) once the profile of the {\it average}
dark matter density is specified. This implies for example that the
Galactic halo signal dominates over the cosmological one. For the
$\kappa=2$ (annihilating dark matter) scenario, however, the picture
changes considerably.  In particular, the clumpiness of the dark
matter distribution strongly affects both the angular distribution
of the emission and the overall flux.

  \end{document}